\documentclass[preprint]{emulateapj}

\def\apj{\rm ApJ}
\def\apjl{\rm ApJL}
\def\apjs{\rm ApJS}
\def\aj{\rm AJ}
\def\mnras{\rm MNRAS}

\def\pasp{\rm PASP}
\def\aap{\rm AAP}

\shorttitle{The Role of ROI in Dark Matter Halos}
\shortauthors{Bellovary et al.}

\begin{document}

\title{The Role of the Radial Orbit Instability in Dark Matter Halo
Formation and Structure}

\author{Jillian M. Bellovary\altaffilmark{1}, Julianne J. Dalcanton\altaffilmark{1}, Arif Babul\altaffilmark{2}, Thomas R. Quinn\altaffilmark{1}, Ryan W. Maas\altaffilmark{1}, Crystal G. Austin\altaffilmark{3}, Liliya L. R. Williams\altaffilmark{3}, and Eric I. Barnes\altaffilmark{4}}

\email{jillian@astro.washington.edu}

\altaffiltext{1}{Department of Astronomy, University of Washington, Seattle, WA}
\altaffiltext{2}{Department of Physics and Astronomy, University of Victoria, Victoria, BC, Canada}
\altaffiltext{3}{Department of Astronomy, University of Minnesota, Minneapolis, MN}
\altaffiltext{4}{Department of Physics, University of Wisconsin -- La Crosse, La Crosse, WI}

\begin{abstract}
For a decade, N--body simulations have revealed a nearly universal
dark matter density profile, which appears to be robust to changes in
the overall density of the universe and the underlying power spectrum.
Despite its universality, the physical origin of this profile has not
yet been well understood.  Semi--analytic models by \citet{Barnes05}
have suggested that the density structure of dark matter halos is
determined by the onset of the radial orbit instability (ROI).  We
have tested this hypothesis using N--body simulations of collapsing
dark matter halos with a variety of initial conditions.  For
dynamically cold initial conditions, the resulting halo structures are
triaxial in shape, due to the mild aspect of the instability.  We
examine how variations in initial velocity dispersion affect the onset
of the instability, and find that an isotropic velocity dispersion can
suppress the ROI entirely, while a purely radial dispersion does not.
The quantity $\sigma^2/v_c^2$ is a criterion for instability, where
regions with $\sigma^2/v_c^2 \lesssim 1$ become triaxial due to the
ROI or other perturbations.  We also find that the radial orbit
instability sets a scale length at which the velocity dispersion
changes rapidly from isotropic to radially anisotropic.  This scale
length is proportional to the radius at which the density profile
changes shape, as is the case in the semi--analytic models; however,
the coefficient of proportionality is different by a factor of
$\sim$2.5.  We conclude that the radial orbit instability is likely to
be a key physical mechanism responsible for the nearly universal
profiles of simulated dark matter halos.
\end{abstract}

\section{Introduction}

The universal density profile of dark matter halos has been widely
observed in cosmological N--body simulations
\citep{Cole96,Moore99,Bullock2001,vdB02,Navarro04}; however, the
physical origins are not yet understood.  Whether formed in an
isolated collapse or by hierarchical merging, dark matter halo density
profiles have a characteristic double-power law shape.  This shape is
usually discussed in terms of the slope of the density profile $\gamma
\equiv d(\log{\rho})/d(\log{r})$.  The canonical halo profile has
$\gamma \sim -1$ in the inner regions and $\gamma \sim -3$ in the
outer regions \citep{Dubinski91,NFW,Navarro97,Huss99,MacMillan06}.
While there are disagreements on the exact values of $\gamma$, the
halo density profiles appear to be similar over decades in radius.
There is also evidence that such profiles may accurately fit the
luminous parts of early--type galaxies as well \citep{Dalcanton01,Merritt05}.

Within dark matter halos, quantities other than density show universal
 profiles as well.  For example, \citet{Taylor01} and later
 \citet{Dehnen05} showed that the phase--space density, represented by
 $\rho/\sigma^3$, has a constant power-law slope of $\alpha \sim 1.9$
 for over two and a half decades in radius in spite of a varying
 density profile (see also \citet{Austin05} for an analytic
 exploration of this behavior).  There is also evidence for a
 universality in the relation between the slope of the density profile
 and the velocity anisotropy.  The anisotropy parameter, $\beta$ is
 defined as

\begin{equation}
\beta \equiv 1 - \sigma_{\phi}^2/2\sigma_r^2,
\end{equation}

\noindent
 where $\sigma_{\phi}$ is the velocity dispersion in the tangential
direction and $\sigma_r$ is the velocity dispersion in the radial
direction.  $\beta$ describes the degree of anisotropy of the velocity
dispersion in a spherical halo, which in simulations ranges from isotropic in
the core ($\beta=0$) to radially anisotropic at the outskirts
($\beta=1$)\citep{Cole96,Carlberg97,Fukushige01}.  \citet{Hansen06}
find a direct correlation between $\beta$ and the density profile
slope for halos with a wide variety of initial conditions
(i.e. isolated collapse, mergers).  \citet{Hanseneveryone06} find
another universality, this time in the velocity distribution function,
for a similar variety of initial conditions.   

The radial orbit instability (ROI) may be the cause of the apparent
universality of these dark matter halo properties.  The ROI occurs in
anisotropic spherical systems composed of particles with predominantly
radial orbits.  The instability arises when particles in precessing
elongated loop orbits experience a torque due to a slight asymmetry.
This torque causes them to lose some angular momentum and move towards
the center of the system.  Particles with small enough angular momenta
become trapped in box orbits, aligning themselves with the growing bar
and reinforcing the initial perturbation.  As a result, the halo
becomes triaxial in shape.  This phenomenon has been examined in
detail by \citet{Dejonghe88} using N--body simulations, by \citet{
Weinberg91} using linear analysis, and by \citet{Huss99} and
\citet{MacMillan06} for the specific case of isolated dark matter halo
collapse.  These groups find that the onset of the ROI corresponds to
a flattening of the central density cusp of the halo, and may be
responsible for the double power-law shape of the halo.  The ROI is
reviewed in Merritt (1999; \S 6.2). \nocite{Merritt99}

Semi-analytic models have also been used to examine the spherical
collapse of dark matter halos, beginning with \citet{Gunn72}.  These
models \citep{Gott75, Gunn77, Bertschinger85} include only radial
motions and produce single power-law density profiles.  When
non--radial motions are included, halo density profiles range from
power-laws \citep{Ryden87} to NFW--like
\citep{Hiotelis02,LeDelliou03,Ascasibar04}.  Recently, however,
\citet{Barnes05} extended the semi-analytic model of
\citet{Williams04} to include a physical representation of the
ROI. The resulting halos have a double power-law density slope similar
to those of N--body simulations.  \citet{Barnes05} concluded that the
density scale length and the anisotropy radius are correlated, and
that the ROI is directly responsible for the shape of the density
profile.  While \citet{Barnes05}'s work is suggestive, the
complexities of a dynamical instability are not easily captured by
analytic or semi--analytic techniques.  We therefore turn to studying
the link between the ROI and halo structure by using high resolution
N--body simulations.

In this paper, we examine in detail the effect of random motions on
the onset of the ROI and the final structure of collapsed dark matter
halos, and attempt to verify the relation between the scale length and
the anisotropy radius reported by \citet{Barnes05}.  We analyze a set
of N--body simulations of isolated, collapsing dark matter halos with
a variety of initial velocity dispersions to study the evolution of
halo properties.  Considering a range of velocity dispersions allows
us to supress the ROI in some cases, helping us to isolate its
physical effects.  In \S2 we describe our simulations and show that
they are robust to resolution and softening effects.  We describe the
properties of the halos in \S3, including the shape and anisotropy
evolution, the effects of velocity dispersion, and the scale-length --
anisotropy radius relation.  We summarize our results in \S4.

\section{Simulations}

We performed simulations using PKDGRAV \citep{Stadel01,Wadsley04}, a
parallel KD Tree gravity solver.  The initial system is an isolated,
spherical halo with a Gaussian radial density profile, mimicking an
overdensity in the early universe.  All particles are given Hubble
flow velocities, such that the halo is initially expanding.  We evolve
the system in physical coordinates for 12.8 Gyr, which was set to
allow 83\% of the mass of the halo to collapse by the present time;
this corresponds to a starting redshift of $\sim12$ according to the
most recent results from WMAP \citep{WMAP3}.

We ran several simulations with comparable density profiles but with a
range of initial velocity dispersions, allowing us to estimate the
threshold at which the ROI will occur. In principle, a dynamically
``warmer'' system should resist the ROI \citep{Merritt85}, since any
perturbations will be washed out by the tangential velocity
dispersions.  We assigned initial random velocities ($\sigma$) to the
particles assuming a Gaussian distribution with a mean of zero; these
numbers are then added to the existing Hubble flow velocities.  The
amplitudes of these initial velocity dispersions were $1\sigma$,
$2\sigma$, and $3\sigma$, where $\sigma$ is 18\% of the circular
velocity at the virial radius of the final system.  Simulations with
a higher velocity dispersion are too warm to collapse at all.  We
also carried out simulations with no velocity dispersion, at both
standard and high resolution.  Finally, we attempted to isolate the
importance of tangential velocity dispersions by carrying out a
simulation with a pure radial initial velocity dispersion of
$3\sigma$.  Table 1 lists the initial value of the ratio $-2T/W$ (see
\S\ref{sec:globalanis}) for each simulation for a more intuitive grasp
of the value of each $\sigma$.

In addition to various velocity dispersions, we ran simulations with a
range of softenings and particle resolutions.  Our highest resolution
simulation has over 2.8 million particles, while the simulations
testing the effects of velocity dispersion have about one fifth of
that number.  These standard-resolution simulations are comparable in
number of particles to \citet{MacMillan06}, but are much larger than
the simulations in \citet{Huss99}, which have $\sim$10,000 particles
within the virial radius of each halo.  The softening, $\epsilon$, was
set to 0.054 times $r_{200}$ of the final collapsed halo, and
is the same for all simulations.  Time steps were set to an accuracy
criterion proportional to $\eta\sqrt{\epsilon/a}$ where $\eta = 0.2$
is the timestep criterion and $a$ is the acceleration of a particle
\citep{Wadsley04}.  We use a force accuracy criterion of 0.55.  A
summary of all simulations is in Table 1.

\begin{deluxetable*}{lrccc}
\tablecolumns{5} 
\tablewidth{0pc}
\tablecaption{Simulation Properties}
\tablehead{
\colhead{Simulation} & \colhead{\# particles } &\colhead{ Velocity Dispersion\tablenotemark{a} } &\colhead{ Softening\tablenotemark{b} } &\colhead{ Initial -2T/W }  }

\startdata
 V0\_H & 2865813 & 0& 1&1.44  \\   
 V0\_M & 573302 & 0& 1&1.45  \\ 
 V1\_M & 573302  & 1& 1&1.49  \\ 
 V2\_M & 573302  & 2& 1& 1.59 \\  
 V3\_M & 573302  & 3& 1&1.64   \\ 
 V3r\_M & 573302  & 3 (radial only) & 1&1.87  \\
 V0\_M\_S1 & 573302 & 0& 2& 1.45 \\  
 V0\_M\_S2 & 573302 & 0& 0.5& 1.45\\  
\enddata
\tablenotetext{a}{in units of $\sigma$, where $\sigma$ is 18\% of the circular velocity of the ``cold'' standard--resolution halo at $z=0$.}
\tablenotetext{b}{in units of 0.054 times $r_{200}$ of the ``cold'' standard--resolution halo at $z=0$.}

\end{deluxetable*}

\subsection{Resolution and Softening Tests}

\begin{figure*}[t]
\begin{center}
\epsscale{0.9}
\plotone{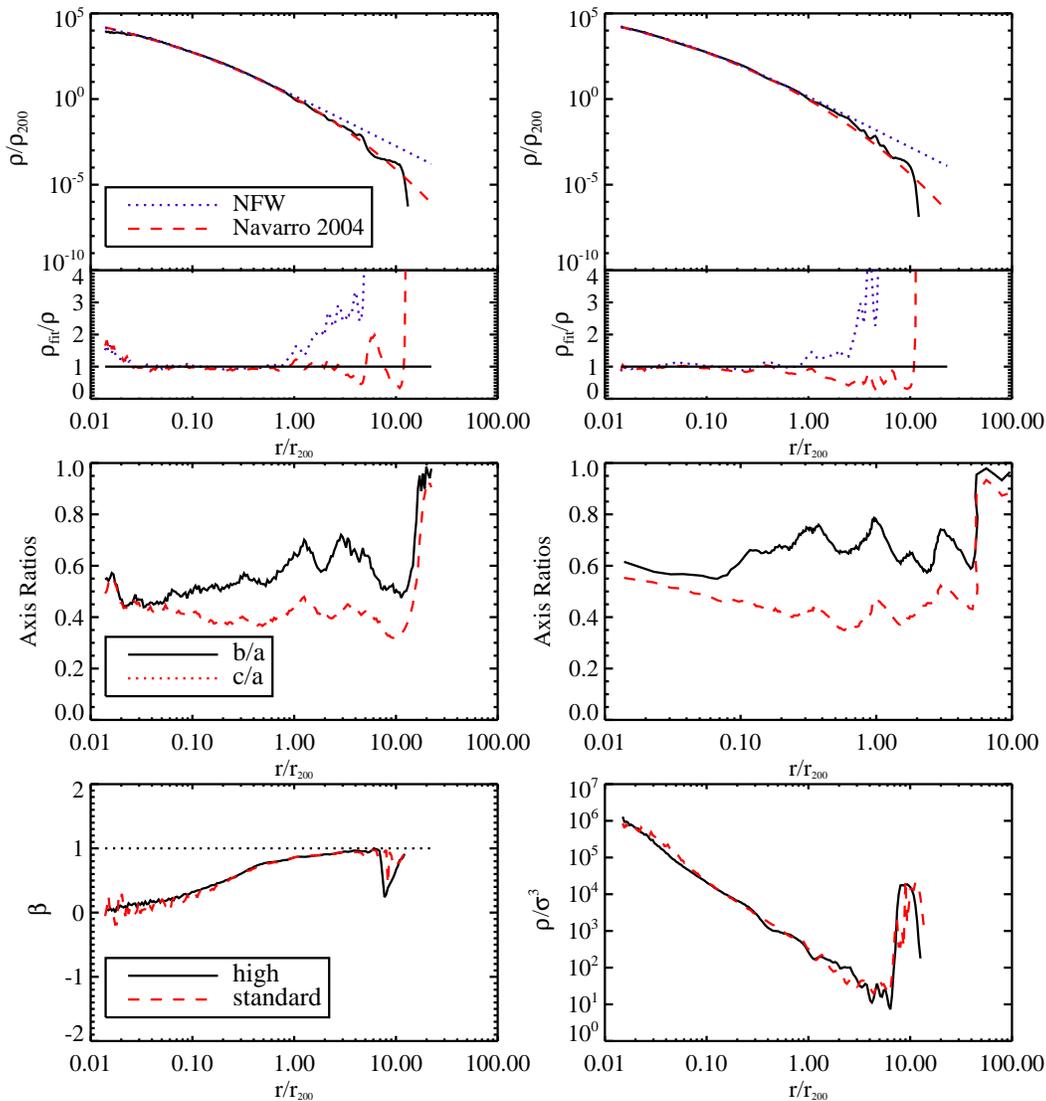}
\caption{Comparison of the standard-- and high--resolution fully
evolved halos.  {\em Top row:} Density profiles of the standard
resolution run (left) and the high resolution run (right).  Fits are
NFW with a concentration of 14.6 (purple dotted line) and the Einasto
profile with $\alpha$ = 0.2 (red dashed line).  The residuals depict
the superior core--fitting in the high--resolution simulation and also
show that the Einasto profile is a good fit to $r \sim 10r_{200}$.
{\em Middle row:} Axis ratio profiles for the standard run (left) and
the high-res run (right). Black line is $b/a$, and red dashed line is
$c/a$.  {\em Bottom left:} Anisotropy profiles for the standard (red
dashed) and high--res (black solid) simulations.  {\em Bottom right:}
phase--space density profiles for the standard (red dashed) and
high--res (black solid) runs.  The standard resolution run shows
slight deviations in density profile at $r < 0.05r_{200}$ and is
slightly less triaxial.  However, the anisotropy and phase--space
density profiles are similar at high and standard resolutions.
\label{fig:restest}}
\end{center}
\end{figure*}

Figure~\ref{fig:restest} shows how the structure of our simulated
halos are affected by particle number (also referred to in the text as
resolution).  The top panels show the density profile for the fully
evolved, $0\sigma$ halo at standard resolution (left) and at high
resolution (right).  Each profile is fit with a NFW profile (purple
dotted line) and a \citet{Navarro04} profile (also known as an Einasto
profile) (red dashed line).  The profiles for the two runs are similar
over a large range in radius. However, as can be seen by the residuals
from the fits, the standard resolution simulation does not adequately
model the inner core of the halo ($r < 0.05r_{200}$, where we define
$r_{200}$ as the virial radius in which the enclosed density is 200
times the critical density of the universe.  There is also a
difference in the innermost cores when looking at the $z=0$ axis ratio
profiles (Fig.~\ref{fig:restest}, middle panels).  The
high--resolution run (right) shows a somewhat more triaxial core,
particularly at the center.  However, the anisotropy and phase--space
density profiles show no significant differences between the two
simulations (Fig.~\ref{fig:restest}, bottom panels, left and right,
respectively).  We analyze these profiles in more detail in \S3.  The
bulk of the difference between the two resolutions is apparent only
within $\sim0.05 \times r_{200}$.  Outside of this radius our
medium--resolution simulations are robust to resolution effects, while
inside this radius the profiles may be less certain, and biased
towards shallow density profiles and rounder shapes.

We have also tested the impact of gravitational softening on our
simulations.  We performed two simulations, identical to the
standard--resolution, $0\sigma$ halo but with softening
increased/decreased by a factor of two.  Decreasing the softening
length does not noticeably change any important quantities (density,
velocity dispersion, axis ratios, phase--space density).  Increasing
the softening length results in a shallower core density profile, due
to the decreased gravitational force at small inter-particle
distances.  The phase--space density profile is similarly affected by
the softening changes; however, the anisotropy profile shows no
difference when the softening lengths are changed.  These effects are
only noticeable within a radius of $0.05 r_{200}$, the same radius
within which we may be affected by resolution limitations ($0.054
r_{200}$).  We conclude that our results are robust beyond a radius of
$0.05 r_{200}$.  This distance corresponds to 10 kpc for a Milky
Way--size halo.

\section{Results}

As the halo collapses, we measure the density profile, velocity
dispersion tensor, and the axis ratios as a function of radius at each
time step.  We expect to see higher central densities and larger
radial velocity dispersions with increasing time.  However, if the ROI
is present we also expect to see deviations from spherical symmetry
develop in regions with large radial velocity dispersions.  These
deviations would be accompanied by increasing tangential velocities due
to the additional torques.

\begin{figure*}[t]
\begin{center}
\epsscale{0.90}
\plotone{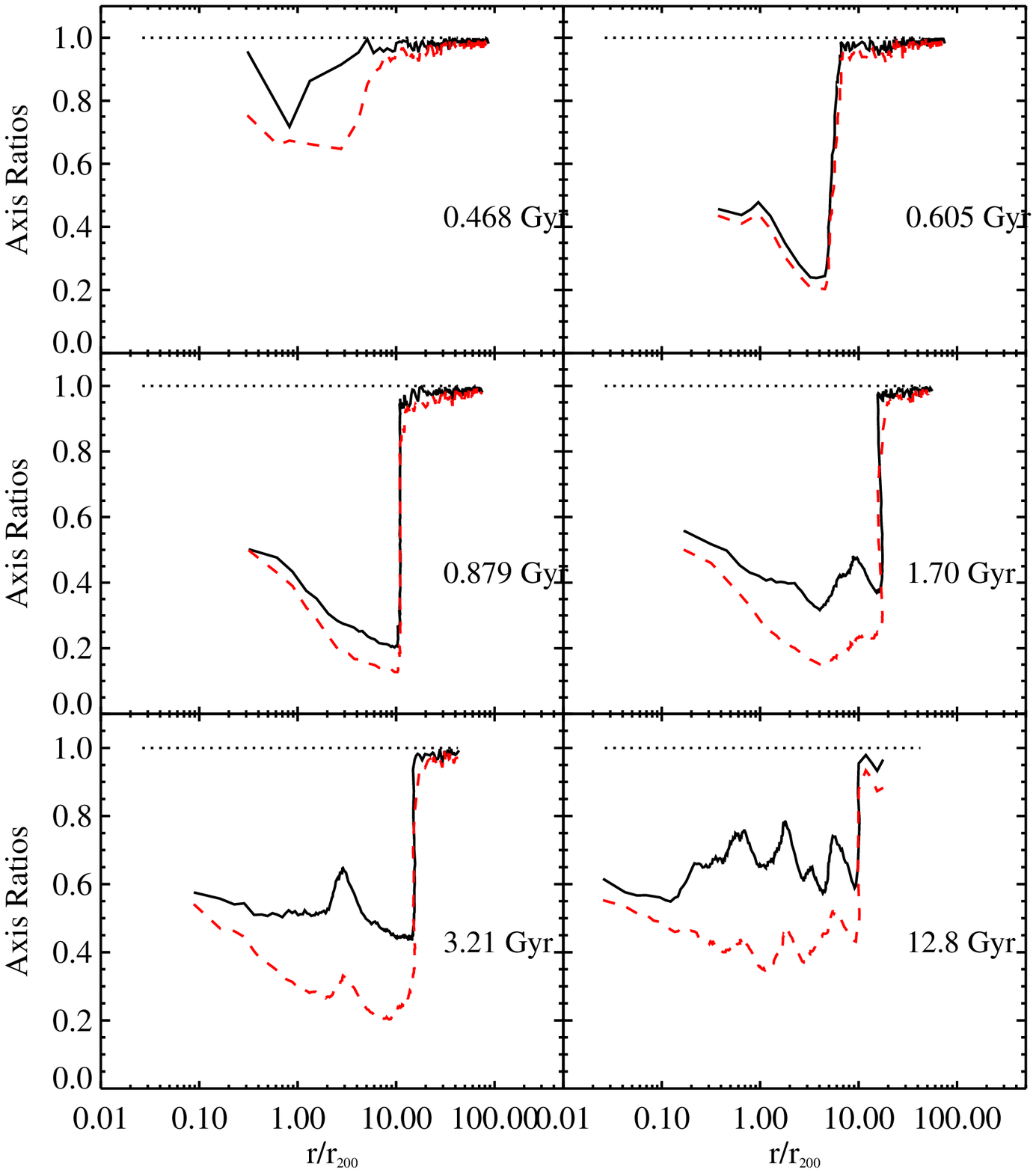}
\caption{Axis ratio profiles of the highest resolution halo at various
times.  Black solid lines are the ratio $b/a$, red dashed lines are
the ratio $c/a$.  The halo profiles quickly become prolate in the
inner regions, with the prolate region expanding to larger radii with
increasing time.  The profiles are normalized to $r_{200}$ at each
timestep.  We interpret the various ``bumps'' in the profiles as being
crossing shells that have not yet reached equilibrium.
\label{fig:nice_ar}}
\end{center}
\end{figure*}

\subsection{Axis Ratios}

The most obvious evidence for the ROI is the growth of triaxiality in
the shape of the halo.  We see this behavior clearly in our
simulations.  We measure axis ratios as a function of radius by
dividing the halo into 196 equal particle-number radial bins based on
particle density, and then calculating the moment-of-inertia tensor
for each bin (see \citet{babul92} for an application of this method).
The square root of the ratios of the principal moments of inertia in
the directions of the three axes of symmetry give the axis ratios of
the particles within each radial bin.  This method more accurately
tracks the radial variations of the axis ratios than methods which
analyze all particles interior to a given radius.  This latter method
is always dominated by the shape of the inner halo, where the majority
of the virialized halo's mass is located, and thus is insensitive to
variations in the axis ratios at large radii.

After 0.6 Gyr, the first shell\footnote{While N--body halos do not
have specific ``shells'' like those in the semi--analytic models, we
use this terminology here to refer to particles at a similar initial radius
and gravitational potential.} reaches the center leading to a large
radial velocity dispersion.  In response, a
triaxial structure develops in the core of the halo.  The resulting
axis ratio profiles can be seen in Figure~\ref{fig:nice_ar}, which
portrays the evolution of the axis ratios $b/a$ and $c/a$ at a range
of timesteps for the high--resolution, zero velocity dispersion halo.
The center of the halo quickly becomes triaxial and remains so
throughout the collapse.  The triaxiality is fairly prolate ($c \sim
0.72b$) and appears as a central bar.  The outer edges of the halo
have yet to collapse, and so remain spherical.  As the collapse
progresses, the triaxial region of the halo grows outwards, and
eventually the majority of the halo settles to an equilibrium $b/a$
ratio of $\sim0.6$.

\begin{figure*}
\begin{center}
\epsscale{0.90}
\plotone{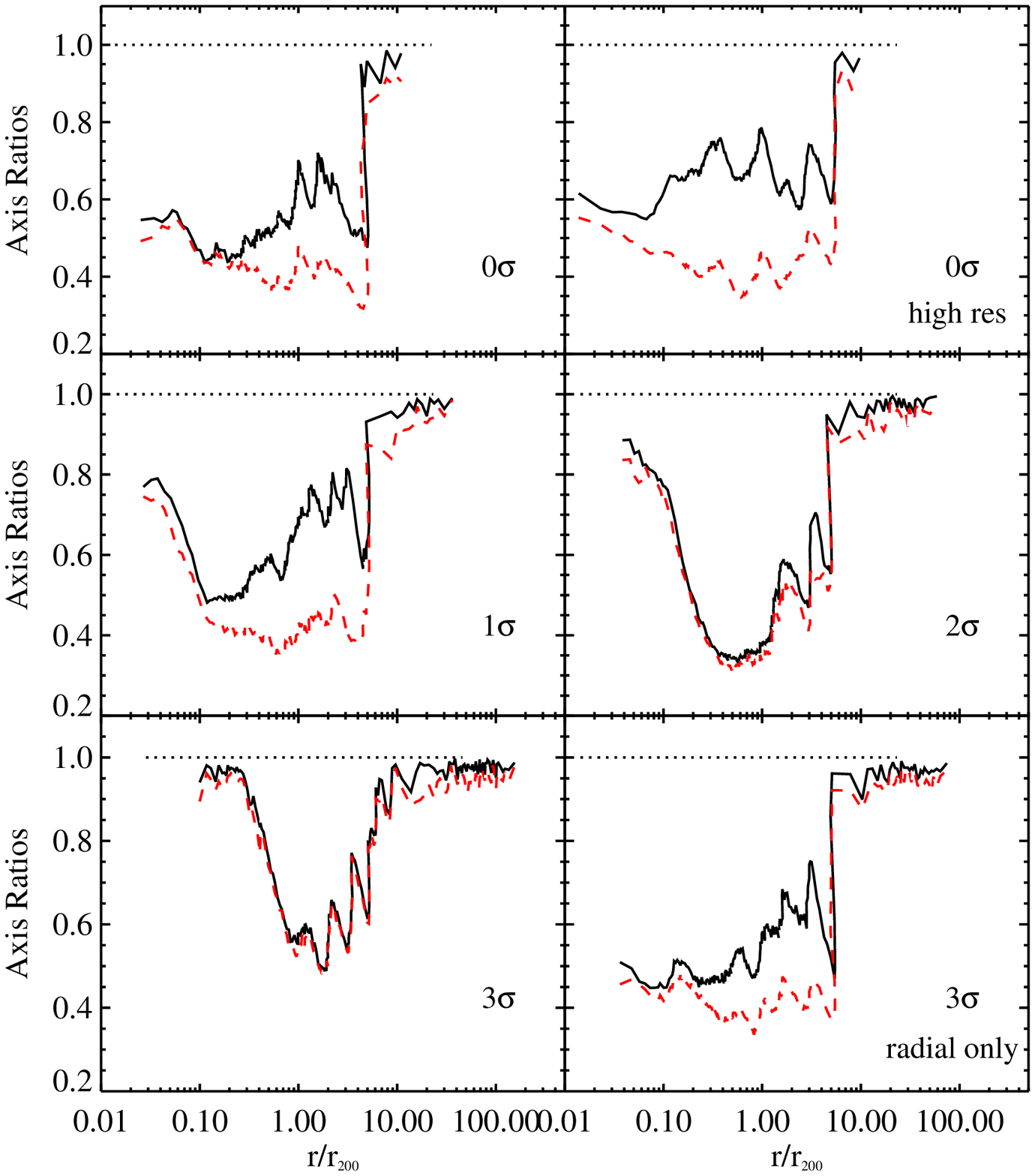}
\caption{The final axis ratio profiles of each simulation.  Black
solid lines are the ratio $b/a$, red dashed lines are the ratio $c/a$.
As the velocity dispersion is increased, the halo becomes increasingly
spherical in the inner regions ($r \lesssim 0.2r_{200}$), suppressing
the ROI.
\label{fig:axis_ratios}}
\end{center}
\end{figure*}

If the formation of the triaxial bar in Figure \ref{fig:nice_ar} is
due to the ROI, then it should be suppressed when the orbits are less
radial.  Our simulations with larger isotropic initial velocity
dispersions would then be expected to show weaker triaxiality.  The
particles in these simulations have larger tangential velocities,
which increases the typical angular momentum of particles' orbits,
suppressing the formation of box orbits, and thus decreasing the
strength of the ROI.  Figure \ref{fig:axis_ratios} shows the resulting
fully evolved axis ratio profiles for a series of simulations with
increasing initial random velocities.  The standard and high
resolution runs with zero velocity dispersion show the central
triaxial signature of the ROI.  The $1\sigma$ halo also shows some of
this structure, but with a rounder core; this halo appears to undergo
a weak form of the ROI.  The 2$\sigma$ and 3$\sigma$ halos, on the
other hand, remain nearly spherical throughout until the end of their
evolution.  Their final shapes are characterized by rather spherical
cores, with the inner axis ratios becoming progressively more
spherical with increasing initial tangential velocities.  A
pre--existing isotropic velocity distribution therefore prevents the
onset of self--gravitating instabilities such as the ROI.  Our results
are supported by \citet{Trenti06}, who find that an isotropic core
acts to stabilize a radially collapsing system against the ROI, even
when the halo is highly anisotropic overall.  Although we do not show
the full time evolution, the halos that develop a strong or moderate
ROI do so quickly, establishing a prolate triaxial core within 1 Gyr
of the onset of collapse.  Thus the triaxial shape indicates the
presence of an instability; a phenomenon such as random scattering of
the particles off of the potential would not create any preferred
shape.

We note that the increase in radial velocity dispersion alone has no
significant effect on the final shape of the halo compared to the
$0\sigma$ collapse.  In the bottom right panel of
Figure~\ref{fig:axis_ratios}, we plot results for a simulation with
large initial radial velocity dispersions, but no tangential
velocities.  Physically, the increased initial velocity dispersion
does little to change the final radial velocity dispersion of the
core, since the latter is dominated by the larger radial motions
produced by the gravitational collapse of the halo as a whole.  Thus,
the resulting axis ratio profile is nearly identical to the case with
no initial velocity dispersion.  The only significant difference is a
slight delay in the collapse of the outer regions.

To determine the physical likelihood of having velocity dispersions as
high as that in the $3\sigma$ halo, we compared our N--body
simulations to the semi--analytic model of \citet{Williams04}, which
treats an evolving perturbation as a series of gravitationally
interacting shells.  The model includes the effects of nearby
substructure, which imparts a velocity dispersion onto the collapsing
shells.  This velocity dispersion increases with time, as the local
substructure becomes more developed and a shell's time of exposure to
the substructure increases.  In contrast, in our N--body simulations
of isolated collapsing halos the velocity dispersions are assigned
only at the outset.

The semi--analytic model shows that a velocity dispersion of $2\sigma$
is given to a shell detaching from the Hubble flow at a redshift of
6.7, while $3\sigma$ is assigned to a shell at $z \sim 4.3$.  These
redshifts correspond to times when the ROI is occurring or has just
occurred in the $0\sigma$ N--body simulations; at these times the
innermost particles' motions are dominated by the gravitational
potential of the halo and will not be strongly affected by external
torques due to substructure.  Subsequent infalling particles will have
increased velocity dispersion due to these effects of substructure;
however, the ROI will have already produced a triaxial core at this
point.  

To summarize, the particles which collapse first and undergo
the ROI may be slightly affected by torques due to substructure, but
these forces are not strong enough to cause the large velocity
dispersions which quench the ROI.  In addition, our $2\sigma$ and
$3\sigma$ halos which do not undergo the ROI are not physically
plausible, since the local substructure cannot impart such large
torques on the innermost particles at the early times at which they
collapse.  This suggests that in a cosmological context, the ROI will
have time to operate in the central regions, before cosmological
torques have a significant effect.

\subsection{Velocity Dispersion Evolution}\label{sec:sigma}

\begin{figure*}
\begin{center}
\epsscale{0.90}
\plotone{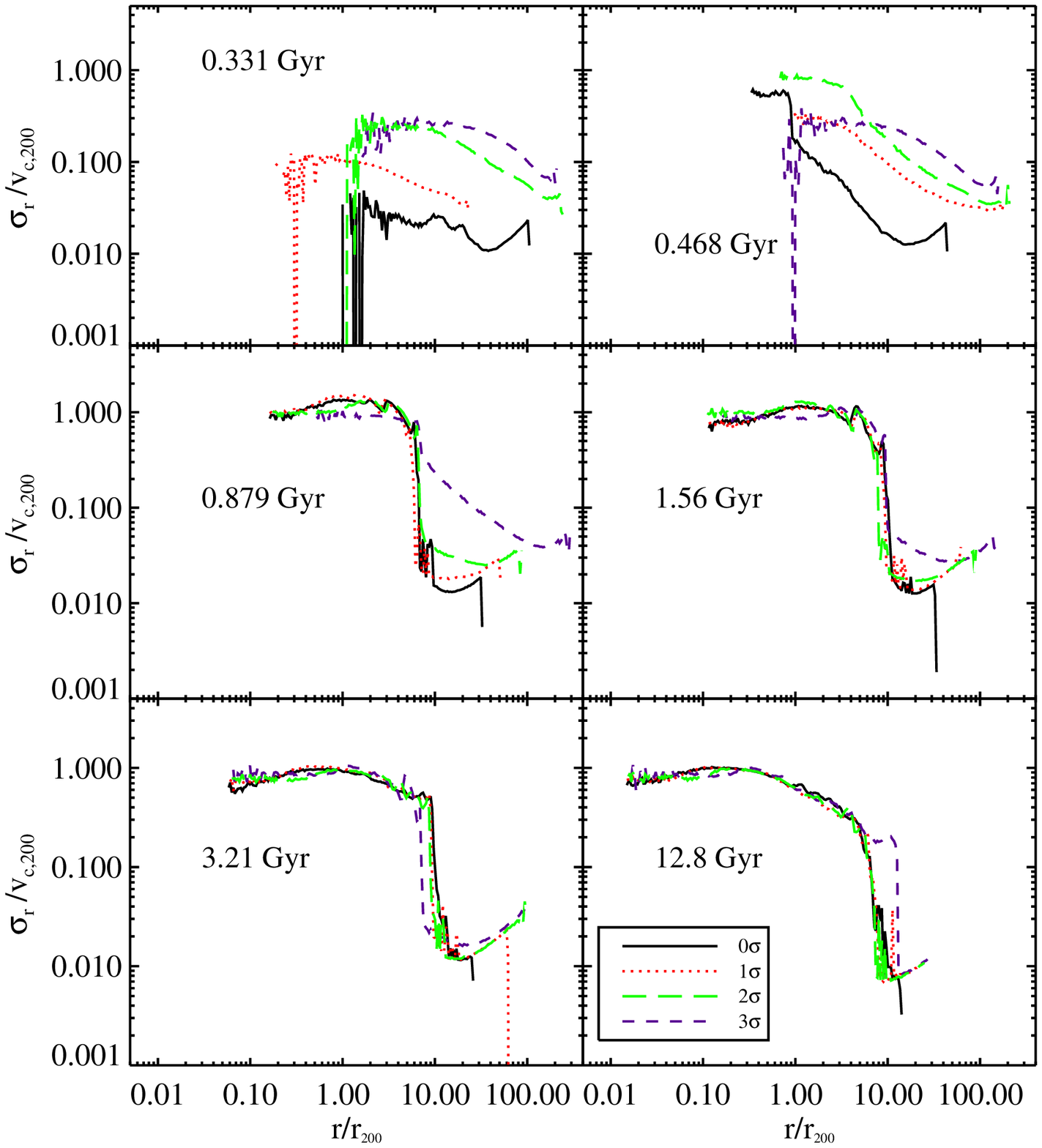}
\caption{The evolution of the radial velocity dispersion profile for
each of the halos.  Simulations represented are no dispersion (black
solid line), $1\sigma$ dispersion (red dotted line), $2\sigma$
dispersion (green long--dashed line), and the $3\sigma$ dispersion
(purple dashed line).  The timestep is labelled on each panel.
Initially, $\sigma_r$ is a small fraction of the final virial
velocity.  As the system virializes, however, $\sigma_r/v_{c,200}$
tends to approach unity.  The virialization clearly occurs from the
inside out.
\label{fig:sr}}
\end{center}
\end{figure*}

If the ROI is responsible for the triaxiality in the inner regions of
low velocity dispersion halos (Figure \ref{fig:axis_ratios}), then we
should expect the onset and radial extent of the triaxiality to be
associated with regions of highly anisotropic radial velocity
dispersions and growing tangential velocities.  For our simulations,
we calculate $\sigma_r$ and $\sigma_\phi$ at each timestep using 196
spherical bins.  Velocity dispersions are determined by subtracting
the average bulk velocity from the velocities in each bin.

\begin{figure*}
\begin{center}
\epsscale{0.90}
\plotone{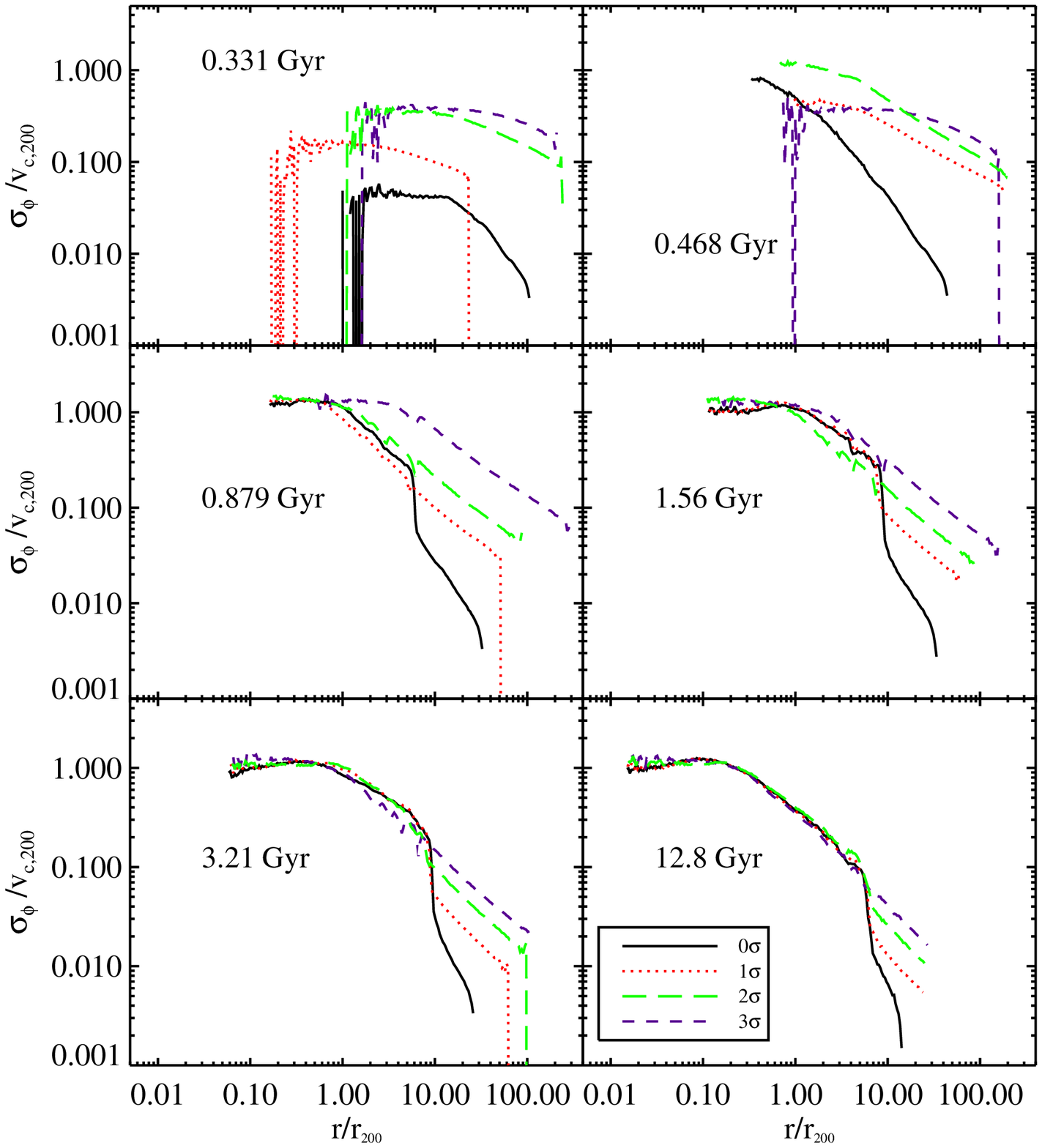}
\caption{The evolution of the tangential velocity dispersion profile for
each of the halos.  Simulations represented are no dispersion (black
solid line), $1\sigma$ dispersion (red dotted line), $2\sigma$
dispersion (green long--dashed line), and the $3\sigma$ dispersion (purple
dashed line).  The timestep is labelled on each panel.
\label{fig:sth}}
\end{center}
\end{figure*}

In Figures \ref{fig:sr} and \ref{fig:sth} we show the radial variation
of the radial and tangential velocity dispersions as a function of
time for our different halos.  The central radial velocity dispersion
of the $0\sigma$ halo undergoes a sharp increase as the innermost regions
collapse into the center of the halo (Figure \ref{fig:sr}, upper
right panel, black solid line).  The radial velocity dispersion
increases outward with time as the halo collapses and material from
larger radii falls inward and virializes.  The $1\sigma$ halo (red
dotted line) undergoes a slower collapse and thus the jump in
$\sigma_r$ is delayed.  The slower collapse is due to the larger
dynamical pressure support provided by the additional initial velocity
dispersion.  However, despite these initial variations in collapse
times, after a few billion years every halo has collapsed and
virialized, and all of the $\sigma_r$ profiles become nearly
indistinguishable.  This likeness is not surprising considering the
similarities in the final density and other profiles (see \S3.4, 3.5).

The tangential velocity dispersion also shows an increase in the
center of each halo early in the evolution (Figure \ref{fig:sth}).  In
the case of the $0\sigma$ halo, this increase is due to the torques
applied by the radial orbit instability.  In the halos which do not
undergo the ROI, $\sigma_{\phi}$ still increases due to the increased
mass in the center of the halo, as the cores grow from material
infalling from large radii.  This increase in the central mass will
cause any orbiting particles to migrate inwards due to the larger
gravitational forces.  However, as the particles move inwards, their
angular velocity must increase to conserve angular momentum.  Thus,
halos with high initial $\sigma_{\phi}$ can {\it also} increase their
tangential velocity dispersion, even when the ROI does not operate.
This effect will also contribute to increasing $\sigma_{\phi}$ in the
$0\sigma$ halo once the ROI has imparted net angular momentum to some
of the particles .  As with the radial velocity dispersion, the final
tangential velocity dispersions are very similar for each halo, even
though the mechanisms to create such profiles are different, and the
resulting halos have different shapes.  The similarity suggests that
after the ROI initially increases $\sigma_{\phi}$, all subsequent
evolution is set by angular momentum conservation during collapse.
Therefore, while the ROI does not uniquely create the observed
universal halo properties such as the NFW density profile, it does
play a crucial role in seeding the initial tangential velocities in
the dynamically ``cold'' collapse case, which then set the subsequent
evolution of the halo.

\begin{figure*}
\begin{center}
\epsscale{0.90}
\plotone{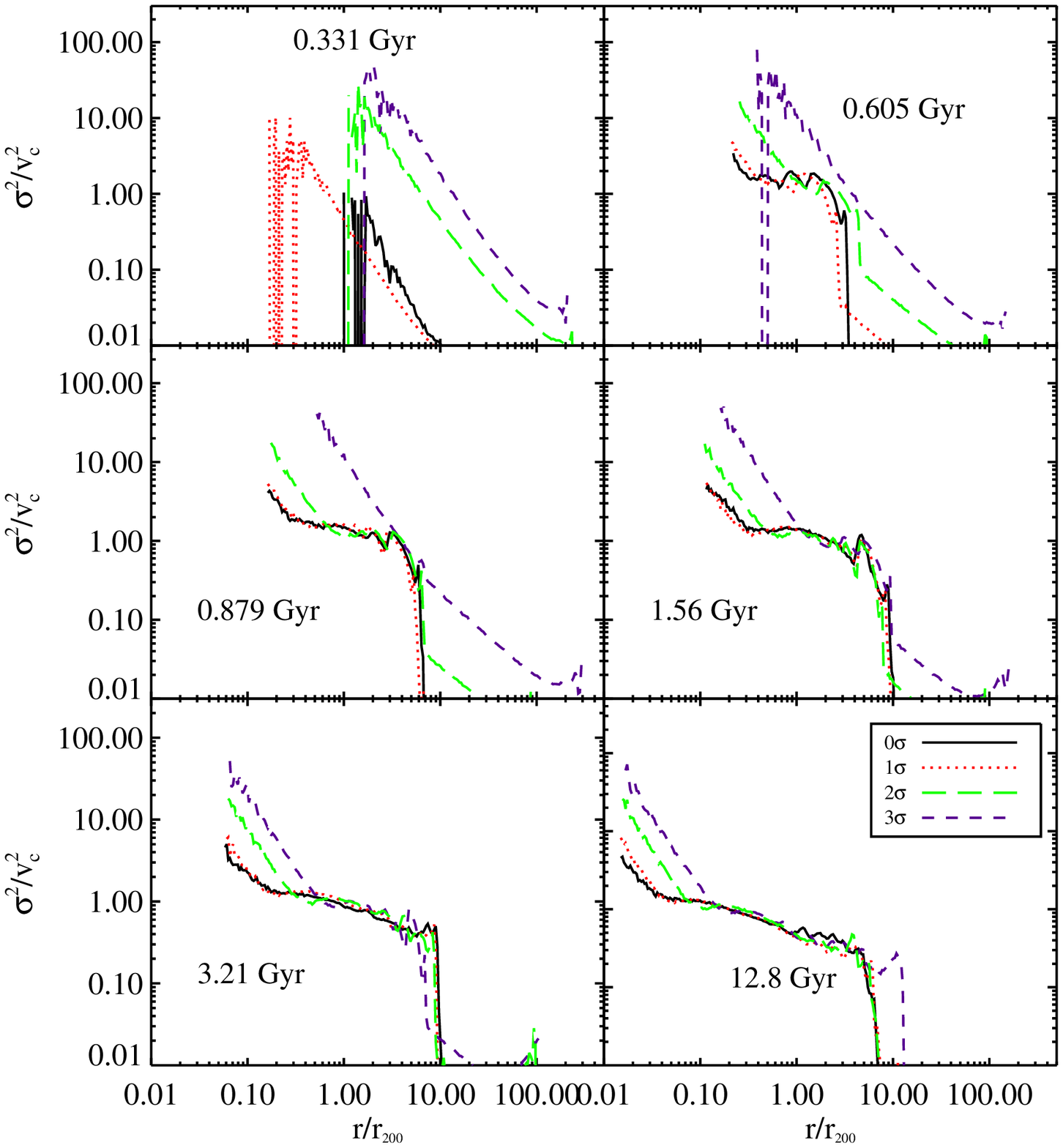}
\caption{The ratio of the square of the velocity dispersion to the
square of the circular velocity with radius, for each simulation at
various times. Simulations represented are no dispersion (black solid
line), $1\sigma$ dispersion (red dotted line), $2\sigma$ dispersion
(green long--dashed line), and the $3\sigma$ dispersion (purple dashed
line).  The timestep is labelled in each panel.  Instabilities grow
when the velocity dispersion is low, and stabilize when the velocity
dispersion becomes comparable to the circular velocity.
\label{fig:ratios}}
\end{center}
\end{figure*}

 Armed with the velocity dispersion evolution shown in Figures
 \ref{fig:sr} and \ref{fig:sth}, we now return to the origin of the
 triaxiality seen for {\it all} halos in Figure \ref{fig:axis_ratios}
 (i.e.\ in the inner regions of the $0\sigma$ and $1\sigma$ halos, and
 the outer regions of the $2\sigma$ and $3\sigma$ halos).  In Figure
 \ref{fig:ratios} we plot the ratio of the square of the velocity
 dispersion over the circular velocity at a given radius.  When the
 velocity dispersions are less than the circular velocity
 (i.e. $\sigma^2/v_c^2 < 1$), halos become unstable to triaxial
 perturbations.  If a slight perturbation occurs, the tangential
 velocity dispersion is not sufficient to wash out the perturbation in
 less than an orbital time.  This is the case at early times for the
 $0\sigma$ and $1\sigma$ halos in the inner regions, as well as in the
 outer regions for the $2\sigma$ and $3\sigma$ halos.  Where this
 mismatch of velocity scales occurs, any triaxial perturbation will
 grow until the ratio of $\sigma^2/v_c^2$ approaches unity.  Comparing
 Figure \ref{fig:ratios} to Figure \ref{fig:axis_ratios} shows that
 indeed the outer prolateness seen in the high--$\sigma$ halos begins
 at the radius where $\sigma^2/v_c^2$ drops below 1.  Interior to this
 point, the velocity dispersions are high enough that any triaxiality
 is smoothed out by random motions before it can be reinforced.  In
 the low--$\sigma$ halos, $\sigma^2/v_c^2$ starts well below 1, but
 the ROI acts quickly to raise the velocity dispersions of the centers
 of the halos, resulting in triaxial cores.

\subsection{Global Anisotropy}\label{sec:globalanis}

The detailed behavior of the velocity dispersions in \S3.2 can be
simplified to a single measure of the velocity structure.  A commonly
used parameter for assessing the velocity structure of a halo is the
global anisotropy:

\begin{equation}
2T_r/T_\phi  \equiv  <\sigma^2_r>/<\sigma^2_\phi>
\end{equation}

\noindent
where $T_r$ and $T_\phi$ are kinetic energies of particles in the
radial and tangential directions, respectively, and the averages are
performed over the entire system.  This quantity will be heavily
weighted by the structure of the inner halo, which contains most of
the halo mass.  The global anisotropy parameter is intended to be used
to describe spherical equilibrium systems, and we note that the most
complete way to compute the anisotropy of an evolving, nonspherical
halo is to make a complete three dimensional map of the velocity
dispersion tensor.  This quantity is difficult to represent, and we
find that when we calculate the global anisotropy parameter it closely
follows the central velocity dispersion tensor.  Thus we use the
quantity $2T_r/T_\phi$ to characterize the typical anisotropy content
at each timestep, even though our halos are evolving and deviate from
spherical symmetry.  Increasing values of the global anisotropy
parameter indicate an increasing prevalence of radial orbits, and for
a spherical equilibrium system, an increased likelihood of undergoing
the ROI.  Though a specific threshold for onset of the ROI has not
been agreed upon, and depends moderately on the initial distribution
function.  Studies have found the threshold to be as low as
$2T_r/T_\phi \sim 1.4$ \citep{BGH} or as high as 2.5 \citep{Merritt85,
Meza97} for equilbrium systems.

\begin{figure}
\begin{center}
\plotone{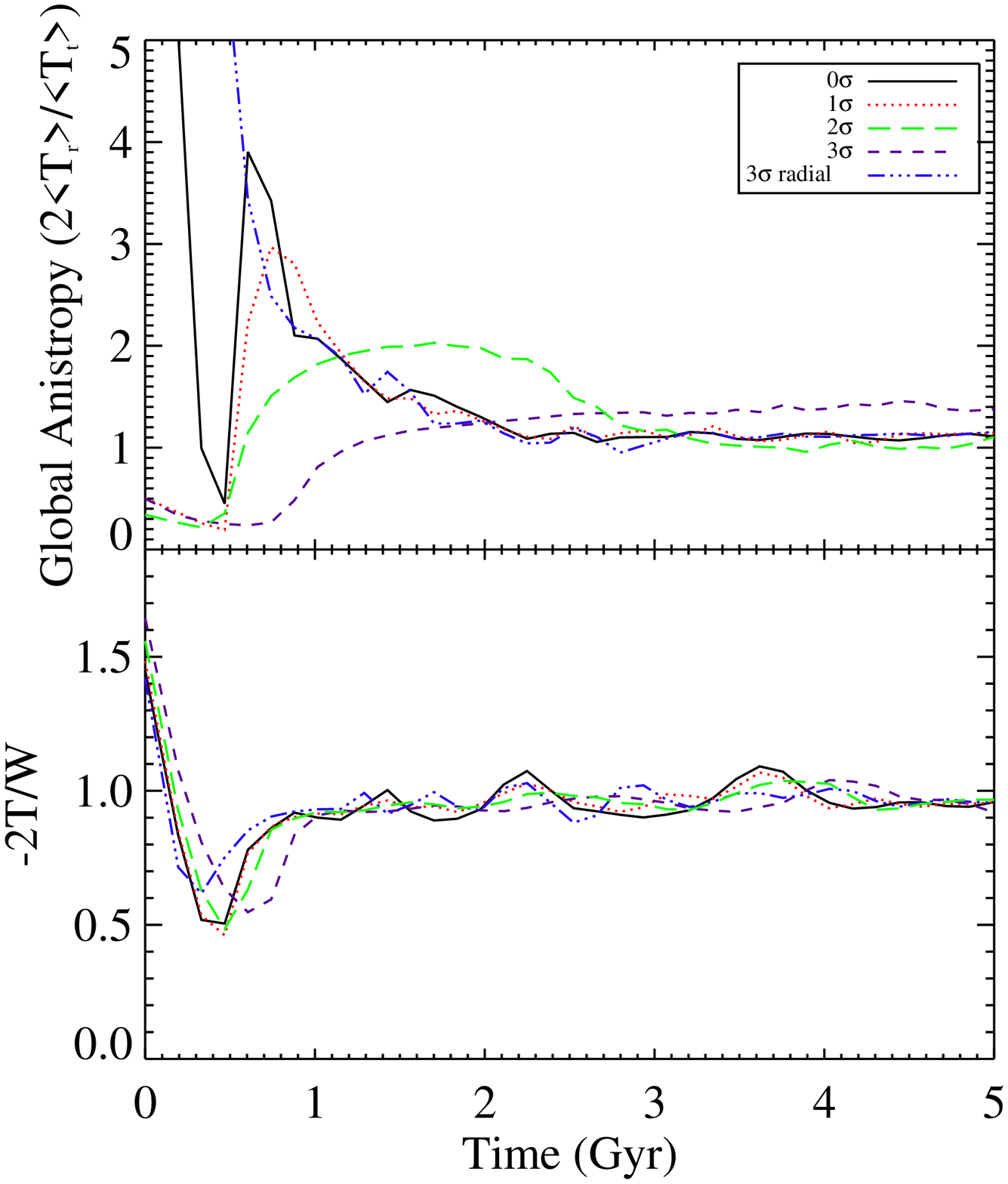}
\caption{({\it Upper Panel}) Global anisotropy of each halo versus
time, showing no dispersion (black solid line), $1\sigma$ dispersion
(red dotted line), $2\sigma$ dispersion (green long--dashed line),
$3\sigma$ dispersion (purple dashed line), and the $3\sigma$ radial
only dispersion (blue dot-dashed line).  The ROI does not occur in
systems with isotropic $\sigma$ of 2 or higher.  ({\it Lower Panel})
Ratio of twice the kinetic energy of the halo to the potential energy
versus time.  The ROI occurs on a different timescale than the
virialization process.
\label{fig:global_anis}}
\end{center}
\end{figure}

At each timestep, the values $<\sigma_r^2>$ and $<\sigma_\phi^2>$ for
the entire halo are found by averaging the squares of the values of
$\sigma_r$ and $\sigma_\phi$ (see \S \ref{sec:sigma}) from every
radial bin normalized by the number of particles in each bin.  The
resulting values of $2T_r/T_\phi$ vs.\ time are presented in the upper
panel of Figure~\ref{fig:global_anis}, for the various initial
velocity dispersions.  At 0.5 Gyr, most halos have $2T_r/T_\phi
\lesssim 0.5$ (the $0\sigma$ and $3\sigma$ radial-only halos have
negligibly small tangential motions at this time, causing the value of
$2T_r/T_\phi$ to be extremely large).  However, the subsequent
evolution of the profiles is quite different.  The low velocity halos
that undergo the ROI show strong increases in $2T_r/T_\phi$, which
peaks at 0.6 Gyr, when the large infall velocities produce large
values of $<\sigma_r^2>$.  In contrast, the higher velocity dispersion
halos (which do not appear to undergo the ROI) show only mild initial
increases in $2T_r/T_\phi$, due to their larger tangential velocities.
For every halo, these increases correspond directly to the time when
each halo deviates noticeably from its original spherical shape.

Following the peak in $2T_r/T_\phi$, the onset of the ROI in the
low-$\sigma$ halos produces a bar which torques the halo particles.
These torques produce larger tangential velocity dispersion,
increasing $<\sigma_{\phi}^2>$ and thus decreasing $2T_r/T_\phi$ to
its final value of $\sim1.2$.  Note that all halos converge to this
final value, regardless of the presence of the ROI.  We plot the time
to 5 Gyr to better portray the evolution of the halos; after this
time, the values of $2T_r/T_\phi$ are fairly constant throughout the
remainder of the simulation.

Given the absence of the ROI triaxiality in the $2\sigma$ and
$3\sigma$ halos, we can use the range between the peak values of the
$1\sigma$ and $2\sigma$ runs to estimate the global anisotropy
threshold for the ROI in our simulations.  This estimate is a fairly
loose one, since the $1\sigma$ halo is not exactly spherical, nor in
equilibrium.  Nonetheless, the resulting value is $ 2.0 < 2T_r/T_\phi
< 2.97$ for our halos, the lower range of which is consistent with
previous estimates for spherical equilibrium models.

The changes in anisotropy described above are not directly associated
with virialization.  The lower panel of Figure~\ref{fig:global_anis}
shows the quantity $-2T/W$ vs. time, where $T$ and $W$ are the sums of
the kinetic and potential energies of each particle in a halo at each
timestep.  This quantity tracks the virialization process of the halo,
which happens on a different timescale than the anisotropy evolution.
Initially, each halo is a diffuse sphere expanding with the Hubble
flow, so there is significant kinetic energy and modest potential
energy.  As the simulation proceeds, the particles reach turnaround
and begin to collapse, at which point the total kinetic energy reaches
a minimum along with the quantity $|2T/W|$.  The ROI then proceeds
directly after the primary collapse of the halo, as can be seen by the
coeval increase in $2T_r/T_\phi$ (isotropization) with the rise of
$|2T/W|$ (virialization) for each halo.  However, isotropization due
to the ROI continues after virialization, as can be seen in the broad
peak in $2T_r/T_\phi$, which continues to evolve after $|2T/W|$ has
reached its final value.  These different timescales emphasize the
distinctness of these two processes.

\subsection{Anisotropy Profile}

Closely related to the global anisotropy is the anisotropy parameter,
$\beta$ (Eqn 1).  The parameter $\beta(r)$ is also defined for a
spherical system, but we use it here as an approximate indication of
the amount of radial motion in the orbits of our halos.  When $\beta =
0$, the velocity dispersions $\sigma_{r}$ and $\sigma_{\phi}$ are
comparable and indicate an isotropic system.  When $\beta = 1$,
$\sigma_{r}$ is much greater than $\sigma_{\phi}$; the orbits in such
regions are dominated by infall, ``shell--crossing'', and in general
highly radial orbits.  We examine $\beta$ calculated in radial shells
to characterize the typical orbits of particles in a given region.
The resulting anisotropy profile $\beta(r)$ can be combined with the
shape information to isolate the dynamical effects taking place.  We
calculate $\beta(r)$ for each timestep using the values of $\sigma_r$
and $\sigma_{\phi}$ described in \S \ref{sec:sigma}.

\begin{figure}
\begin{center}
\plotone{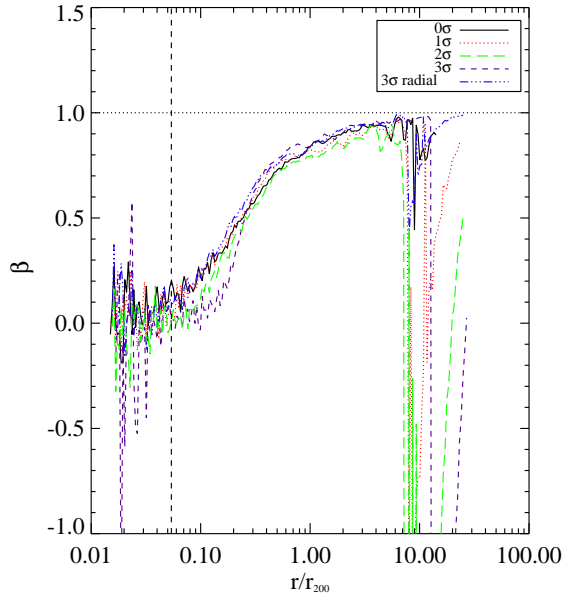}
\caption{Anisotropy profiles for halos with no velocity dispersion
(black solid line), $1\sigma$ dispersion (red dotted line), $2\sigma$
dispersion (green long--dashed line), $3\sigma$ dispersion (purple
dashed line), and the $3\sigma$ radial only dispersion (blue
dot-dashed line).The horizontal black dotted line marks $\beta = 1$
for reference, and the vertical black dashed line marks the softening
length.  All halos develop isotropic cores and radially--dominated
outer regions, regardless of initial velocity dispersions.
\label{fig:beta_sigmas}}
\end{center}
\end{figure}

  Our halos all settle into a common final anisotropy profile,
characterized by a roughly isotropic core surrounded by an
anisotropic, radially--dominated outer region, as shown in
Figure~\ref{fig:beta_sigmas} and as noted in several other papers
including \citet{Barnes05}. However, we find that this profile occurs
regardless of initial velocity dispersion, and whether or not the ROI
has taken place.  We can explain this behavior as follows: the
anisotropy of the outer regions is expected, since material at large
radii predominately consists of inward and outward moving particles
that began with a high gravitational potential, hence having large
characteristic radial velocities.  In the center, velocities are
isotropic for one of two reasons: either the ROI has occurred,
transferring radial motions to tangential, or the motions are
dominated by the initial isotropic velocity dispersion imposed by the
initial conditions.  In the latter case, any increase in radial
velocity dispersion due to infalling shells is counteracted by the
increase in tangential velocities as particles attempt to conserve
angular momentum in response to being pulled inward by the infalling
material.  Therefore, both the large and small radius behavior suggest
that the behavior of $\beta(r)$ increasing from 0 to 1 should be quite
typical, as we indeed find.

Although the overall shape of $\beta(r)=0\rightarrow1$ is similar among
all halos, we do see a slight signature of the ROI in the steepness
with which $\beta(r)$ increases.  The higher--$\sigma$ halos have
noticeably steeper transitions from $\beta(r) = 0 \rightarrow 1$, which
could possibly be a result of the initial velocity dispersion
dominating the core isotropy rather than the ROI.

It is important to point out that a triaxial halo core cannot be
purely isotropic, though it appears so in
Figure~\ref{fig:beta_sigmas}.  This is due to our use of spherical
radial bins to analyze a non-spherical system, and the fact that the
standard-resolution halos do not have enough particles in the central
region to properly resolve the orbital motions.  An examination the
Cartesian velocity dispersion tensor of the central regions confirms
that the core is nearly but not exactly isotropic.  Our use of
$\beta(r)$ as a measure of anisotropy is still very useful, however,
as it effectively tracks large--scale changes in anisotropy with
radius,as can be seen from the figures.

\begin{figure}
\begin{center}
\plotone{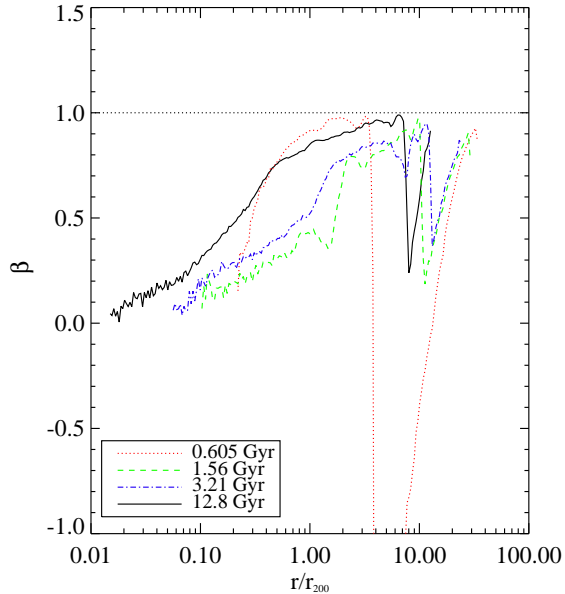}
\caption{Evolution of the anisotropy profile with time for the
high--resolution, $0\sigma$ halo. Red dotted line is $t=0.605$ Gyr,
green dashed line is $t=1.56$ Gyr, blue dot-dashed line is $t=3.21$
Gyr, black solid line is $t=12.8$ Gyr.  The black dotted line marks
$\beta = 1$ for reference.  Each profile is normalized to the virial
radius at that timestep.  The characteristic shape of $\beta(r)$ forms
early and is maintained throughout the halo's evolution.
\label{fig:beta_red}}
\end{center}
\end{figure}

We track the time evolution of the anisotropy profile in
Figure~\ref{fig:beta_red} for the high--resolution $0\sigma$ halo.
The characteristic shape of the profile develops very early on as
particles at small radii collapse and undergo mixing.  At the first
timestep shown, the ROI has just begun, causing the innermost
particles to enter box orbits, while the rest of the halo is still
dominated by radial infall (red dotted line).  The near-isotropic core
develops quickly and grows as more particles reach the center; by 1.6
Gyr, the ROI has operated on much of the center of the halo (green
dashed line).  The outermost regions consist of radially infalling
material throughout the evolution of the halo.  Note that the
high-resolution halo has enough particles to resolve the orbital
motions at the core.  The $\beta(r)$ profile tends toward zero at
small radii but an anisotropic component remains, due to the
triaxiality of the halo.

The time evolution of the anisotropy profiles of the higher--$\sigma$
 halo is comparable to those of the ``cold'' halo, in that the cores
 of the halos become isotropic at early times.  However, while these
 halos appear to develop similarly, the core isotropy comes about
 differently for each halo; the higher--$\sigma$ halos have
 pre--existing isotropic orbits in the central region, preventing
 instabilities during collapse and maintaining a central isotropic
 core.  In contrast, the 0$\sigma$ halo develops its isotropic
 core via the ROI.  Therefore, the characteristic shape of the
 anisotropy profile is independent of the shape of the halo, and can
 be formed either when there is an existing isotropic velocity
 dispersion, or via such a mechanism as the ROI if the halo is
 dynamically cold.  

This finding explains the results of \citet{Lu06},
 who find that the universal $\gamma\sim-1$ inner halo density slope
 is formed for a variety of initial velocity dispersion distributions
 (i.e. radial, isotropic), particularly during an early ``fast
 accretion'' phase.  Our Gaussian initial conditions are such that
 $\sim 90\%$ of the halo mass has a freefall time within 50\% of the
 median.  Thus our halos are likely to fall in the ``fast accretion''
 regime of \citet{Lu06}.  Note, however, that unlike \citet{Lu06} we
 do not need to posit an additional source of isotropization for a
 core to be produced.  The necessary isotropization occurs naturally
 either through the ROI or the conservation of angular momentum.

\subsection{Density and Phase--Space Density Profiles}

\begin{figure}
\begin{center}
\plotone{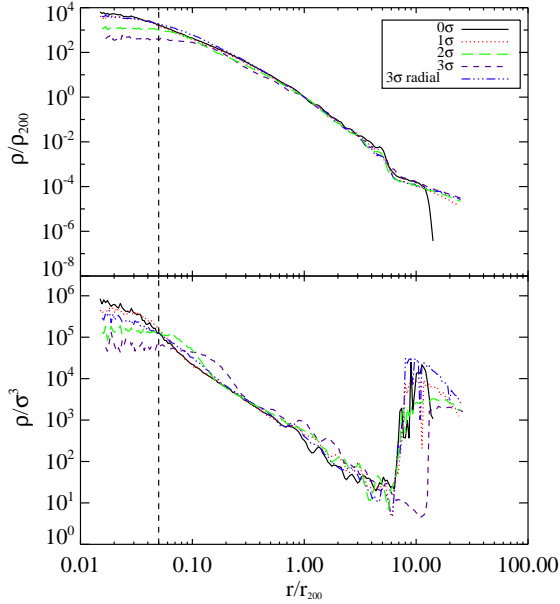}
\caption{ Fully evolved density and phase--space density profiles for
halos with no velocity dispersion (black solid line), $1\sigma$
dispersion (red dotted line), $2\sigma$ dispersion (green long--dashed
line), $3\sigma$ dispersion (purple dashed line), and the $3\sigma$
radial only dispersion (blue dot-dashed line).  The vertical dashed
line marks the radius below which softening and resolution begin to
affect the profiles.  All of the profiles are very similar, except that
halos with higher velocity dispersion have shallower cores.
\label{fig:den_psd_sigmas}}
\end{center}
\end{figure}

Halos with various velocity dispersions show similar structural
properties, whether or not they undergo the ROI.
Figure~\ref{fig:den_psd_sigmas} plots the fully evolved density and
phase--space density profiles of each standard--resolution halo.
There is a noticeably shallower core in both the density and
phase--space density profiles with higher $\sigma$. This effect was
first reported by \citet{Merritt85} and is not surprising,
considering the higher initial angular momenta characteristic of
systems with large $\sigma_\phi$.  It is also important to recall from
our tests in \S2 that these medium--resolution simulations may not
completely capture the properties of the halo core.  So, while these
results certainly make intuitive sense, high--resolution simulations
are needed for confirmation within $0.05r_{200}$.

\begin{figure}
\begin{center}
\plotone{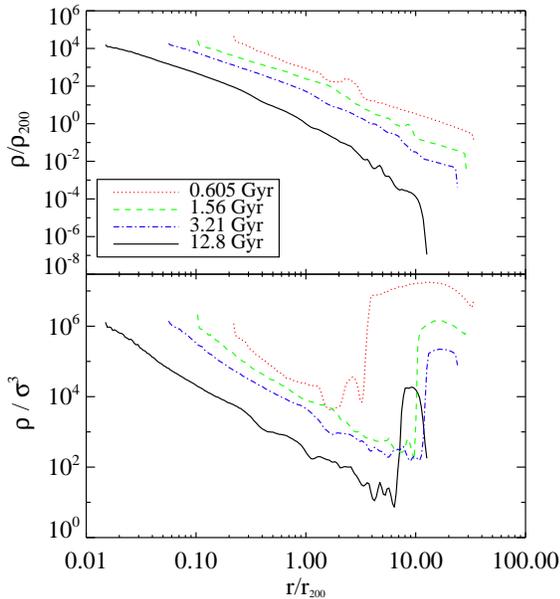}
\caption{Evolution of the density and phase--space density profiles
with time for the high--resolution, $0\sigma$ halo. Red dotted line is
$t=0.605$ Gyr, green dashed line is $t=1.56$ Gyr, blue dot-dashed line
is $t=3.21$ Gyr, black solid line is $t=12.8$ Gyr.  All density curves are normalized to $\rho_{200}$ at $t=12.8$ Gyr to better show the density profile evolution. The profile shapes
are set early on and evolve radially outwards with time.
\label{fig:den_psd_red}}
\end{center}
\end{figure}

The evolution of the density and phase--space density profiles with
time are shown in Figure~\ref{fig:den_psd_red} for the
high--resolution, zero dispersion halo.  Each profile is normalized to
the virial radius at each corresponding timestep.  The basic shape of
each profile is formed after 1.5 Gyr of evolution and grows outward
with radius as the halo collapses.  The outermost radii have yet to
collapse at the present time.  The final density profile (top panel)
is fit well out to $10r_{200}$ by an Einasto profile with $\alpha$ =
0.2.  The NFW profile with $c=8.0$ gives a poorer fit at the outer
edges (see Fig. 1, top panels).

 Our halos exhibit a power-law in the phase--space density
(represented by $\rho/\sigma^3$) with a slope of -1.7 that is
relatively constant with time.  The final halo power-law persists over
$\sim2.5$ decades in radius.  The core phase--space density begins to
develop its power-law structure at early times as well
(Figure~\ref{fig:den_psd_red}, bottom panel).  These results are consistent with other simulations and do not seem to be affected by the ROI.

\subsection{Scale Lengths}

\citet{Barnes05} developed a semi--analytic model of collapsing
halos which includes a physical representation of the radial orbit
instability.  This model varies the initial perturbation velocities of
the collapsing halo in such a way that an isotropic core is produced,
with orbits becoming more anisotropic with increasing radius.  This
model produces realistic NFW density profiles and power--law
phase--space density profiles for a series of halos.  In addition, the
model shows a direct correlation between the anisotropy radius
($r_a$), defined as the radius where $\beta$ = 0.5, and the density
scale length ($r_s$, here defined as the radius where $\gamma$ = -2).
This correlation is assumed to be a result of the ROI acting to
flatten the density cusp by torquing the orbits of the inner halo
particles and forming an isotropic core.

\begin{figure}
\begin{center}
\plotone{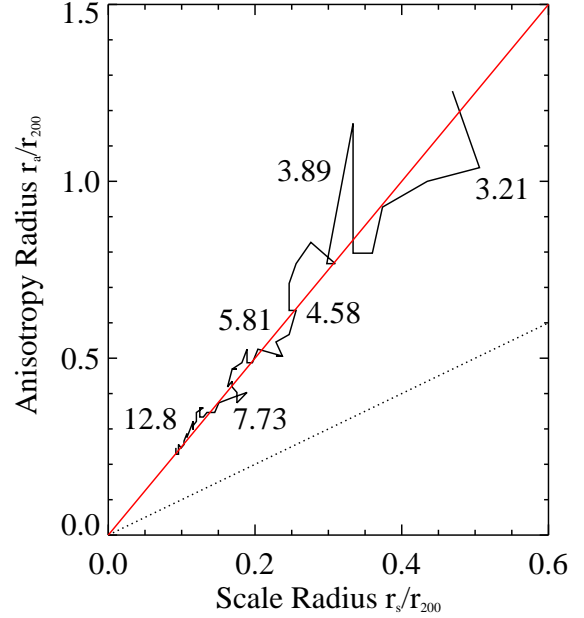}
\caption{Anisotropy radius ($r_{a}/r_{200}$) versus density scale
radius ($r_{s}/r_{200}$) for the standard resolution, $0\sigma$ halo.
These radii appear to be correlated with a slope of 2.5 (red line).
The semi-analytic model of \citet{Barnes05} gives a slope of $\sim 1$
(dotted line). Numbers on the plot indicate the time in Gyr at that
particular point.
\label{fig:rad_rad}}
\end{center}
\end{figure}

While we showed in \S 3.3 that the characteristic anisotropy profile is
not always due to the ROI, we still expect a correlation between $r_a$
and $r_s$.  Within $r_a$, tangential velocities are large, and the
resulting high angular momenta should keep particles from penetrating
the very center of the halo.  This angular momentum barrier should
lead to a flatter core whose size $r_s$ is physically related to the
size of the isotropic region.

To test the conclusion presented in \citet{Barnes05}, in
Figure~\ref{fig:rad_rad} we plot the relation between $r_a$ and $r_s$
as a function of time for the standard--resolution, zero--dispersion
halo.  The numbers on the figure indicate the time in Gyr at that
particular point.  While our N--body simulations do not confirm that
the radii are equal, the two scale lengths are clearly proportional to
each other.  While there is considerable scatter, the points gravitate
toward a line with slope $\sim 2.5$, suggesting that the anisotropy
radius and the scale radius have some connection in their development.
One reason for the difference in the slope of our relation versus that
of \citet{Barnes05} could be their assumption that the ROI does not
change the energies of the particles, only the angular momenta.  Our
N-Body simulations allow for both of these quantities to change,
allowing for a more accurate portrayal of the ROI.  Thus the
possibility remains that the ROI is the mechanism that produces the
characteristic anisotropy profile for a ``cold'' collapsing system.

\section{Conclusions}

We have conducted N--body simulations of a variety of isolated,
collapsing dark matter halos to investigate the role of the
radial orbit instability on the final halo structure.  The ROI plays a
role in determining the eventual shape, density profile, and
phase--space density profile of a halo.  The ROI is suppressed in
halos with isotropic initial velocity dispersions, but halos with
 purely radial velocity dispersions appear to undergo the instability. 

Many numerical simulations have produced triaxial dark matter halos
\citep{Frenk88,Warren92, Thomas98, Jing02}.  Observational evidence
for triaxial halos is also appearing, from low surface brightness and
dwarf galaxies \citep{Bureau99, Simon05, Hayashi06} and from galaxy
clusters \citep{Oguri03, Lee04}. Such evidence for the existence of
triaxial halos strengthens our hypothesis that dynamically cold halos
undergo the ROI during collapse and therefore obtain a triaxial shape.
A warmer system would not become unstable and would have a final shape
that is too spherical to explain the observed structures.  While the
linear overdensities in the early universe are themselves triaxial,
not spherical \citep{BBKS}, and hence also contribute to the
triaxiality of the resulting halos, the ROI significantly affects the
inner, observable parts of the halo.

The characteristic anisotropy profile seen in all of our halos, and in
others in the literature, is independent of the ROI, and appears to be
a result of the general process of halo collapse.  However, in cases
where the ROI does occur, it is likely to be the cause of such
profiles.  In these cases, there is a link between the density scale
radius and the anisotropy radius, indicating a causal relation between
the two.  Further semi--analytic and N--Body simulations may help to
clarify this potential connection.

While simulations of isolated collapsing halos are useful for trying
to understand the effects of the ROI, it is well known that in
realistic cosmological settings, halos form hierarchically.  Their
evolution is marked by periods of gentle accretion, typically in the
form of minor mergers, occasionally punctuated by major mergers.  To
understand the role and relevance of the ROI in halos evolving in this
way, we are presently in the process of carrying out and analyzing
simulations of halos subject to controlled major and minor mergers
(cf. \citep{Poole07}).  At this early stage, we are finding that
the complex dynamics at play during major mergers makes it difficult
to ascertain straightforwardly whether the ROI plays a significant
role during the subsequent relaxation process; a priori, we would
speculate that it does not.  However, there are indications that the
ROI is operational during periods of quiescent accretion, by acting
upon nearly radial tidal streams associated with disrupting subhalos,
which become isotropic.  We speculate that the weak tides induced by
these weak structures in fact both seeds and reinforces the
instability.  

\acknowledgements JMB thanks the NSF Graduate Research Fellowship
Program.  JMB, LLRW, AB, JJD, CGA, EIB, and RWM would like to
acknowledge NSF grant AST--0307604.  JJD was also supported by a
Wyckoff Faculty Fellowship at the University of Washington.  AB
acknowledges support from NSERC (Canada)'s Discovery Grant Program as
well as the Leverhulme Trust for awarding him a Leverhulme Visiting
Professorship.  Computing resources were made possible through a grant
from the Student Technology Fee of the University of Washington.  The
authors thank Andrew West for useful insights and the anonymous
referee for helpful suggestions which improved the paper.

\end{document}